\pdfoutput=1

\documentclass[11pt]{article}

\usepackage[final]{acl}

\usepackage{times}
\usepackage{latexsym}

\usepackage[T1]{fontenc}

\usepackage[utf8]{inputenc}

\usepackage{microtype}

\usepackage{inconsolata}

\usepackage{graphicx}

\graphicspath{{fig/}{./}}

\usepackage{balance} 
\usepackage{bbold}
\usepackage{enumitem}
\usepackage{multirow}
\usepackage{subcaption}
\usepackage{xcolor}
\usepackage{xspace}
\usepackage[export]{adjustbox}
\usepackage{booktabs}
\usepackage{tcolorbox}
\tcbuselibrary{breakable}
\tcbset{breakable}
\usepackage{tikz}

\newcommand{\tool}[0]{\textsf{{AIPOM}}\xspace}
\newcommand*\blackcircled[1]{\tikz[baseline=(char.base)]{
            \node[shape=circle,fill,inner sep=1pt] (char) {\textcolor{white}{#1}};}}
\newcommand*\circled[1]{\tikz[baseline=(char.base)]{
            \node[shape=circle,draw,inner sep=1pt] (char) {#1};}}

\newtcolorbox{promptbox}[2][]{width=\linewidth,
boxsep=2pt,left=8pt,right=7pt,top=5pt,bottom=5pt,
fontupper=\ttfamily,fontlower=\ttfamily,
fonttitle=\hypersetup{linkcolor=white,urlcolor=white},
title={#2},
label={#1}
}

\title{\tool: Agent-aware Interactive Planning for Multi-Agent Systems}

 \author{Hannah Kim, Kushan Mitra, Chen Shen, Dan Zhang, Estevam Hruschka \\
         Megagon Labs \\ 
         \texttt{\{hannah, kushan, chen\_s, dan\_z, estevam\}@megagon.ai}}

\makeatletter
\AtBeginDocument{
\let\@oldmaketitle\@maketitle
\renewcommand{\@maketitle}{\@oldmaketitle
  \vspace{-27pt}
  \includegraphics[width=\linewidth]{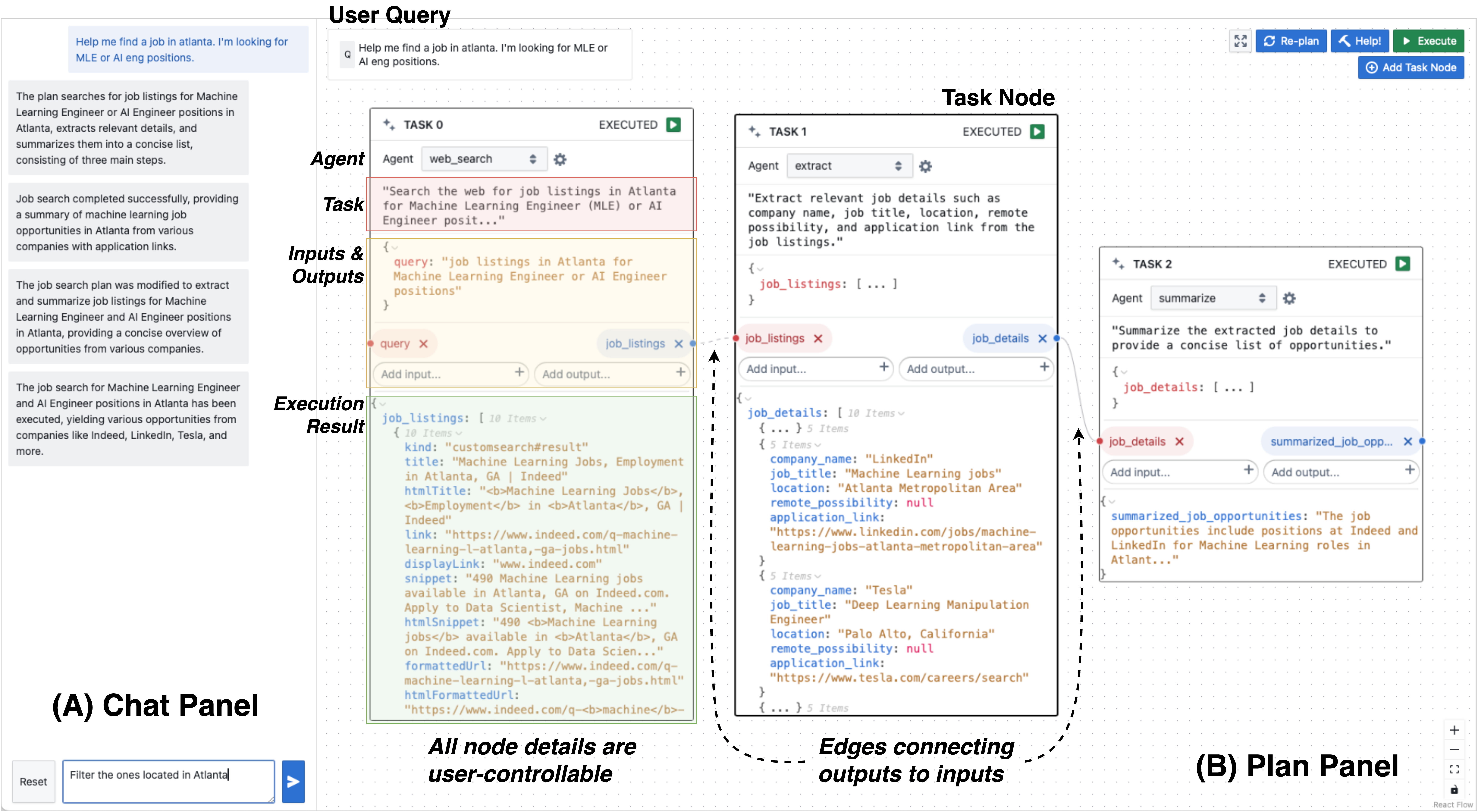}
  \vspace{-18pt}
  \captionof{figure}{
    \tool enables transparent and controllable planning in multi-agent workflows through conversational and graphical interfaces that support human–LLM collaboration.
    \textbf{(A) The Chat Panel} allows users to define or update the planning goal, provide high-level feedback, and receive updates or explanations.
    \textbf{(B) The Plan Panel} displays the generated plan as an editable graph, enabling users to directly manipulate task nodes, agent assignments, data flow, and execution outputs.
  }
  \label{fig:teaser}
  \vspace{17pt}
 }
}
\makeatother

\begin{document}
\maketitle

\begin{abstract}
Large language models (LLMs) are being increasingly used for planning in orchestrated multi-agent systems. However, existing LLM-based approaches often fall short of human expectations and, critically, lack effective mechanisms for users to inspect, understand, and control their behaviors. These limitations call for enhanced transparency, controllability, and human oversight. To address this, we introduce \tool, a system supporting human-in-the-loop planning through conversational and graph-based interfaces. \tool enables users to transparently inspect, refine, and collaboratively guide LLM-generated plans, significantly enhancing user control and trust in multi-agent workflows.
Our code and demo video are available at \url{https://github.com/megagonlabs/aipom}.
\end{abstract}

\section{Introduction}
\label{sec:intro}
Orchestrated Multi-Agent Systems (OMAS) have emerged as a powerful framework for handling complex tasks across diverse domains~\cite{kandogan2024blueprintarchitecturecompoundai, compound-ai-blog}. These systems consist of multiple specialized agents, each responsible for performing specific subtasks upon request. The agents are systematically orchestrated, with their outputs propagating through successive agents to collaboratively resolve a given task.  Recently, these modular workflows have been enhanced by the integration of large language models (LLMs), external tools, and domain-specific models, leading to improved performance and adaptability in tackling complex, real-world tasks \cite{schick2023toolformer, chen2024are}.

A key component of OMAS is planning, i.e., the process of breaking down high-level goals into structured sequences of subtasks and assigning them to appropriate agents. LLMs are increasingly being used for planning~\cite{huang2022language,wang2023describe,10161317}, owing to their ability to perform complex reasoning, generalize across domains, leverage world knowledge, reflect on their own planning decisions, and operate directly through natural language \cite{renze2024self,zhang2025learning}. These capabilities make LLMs well-suited for orchestrating multi-agent interactions without task-specific training. 

Despite these strengths, LLM-based planning presents several challenges. First, in domain-specific or high-stakes scenarios, LLMs may generate outputs that are inaccurate, incomplete, or misaligned with expert knowledge~\cite{valmeekam2023on,huang2024understandingplanningllmagents}. Second, in many OMAS settings, users are presented only with the final output of the system, without visibility into the underlying plan structure or the intermediate outputs produced by agents. This lack of transparency makes it difficult to understand, verify, and trust the system’s behavior. Finally, these systems are typically accessed through chat interfaces, which offer limited controllability and make it difficult for users to inspect, refine, or debug plans at a granular level. These limitations make human oversight not only necessary but central to the planning phase, underscoring the need for interfaces that allow users to actively engage with and guide the planning and execution processes to ensure outcomes align with their intentions~\cite{EU}.

To address these challenges, we present \tool (Agent-aware Interactive Planning for Orchestrated Multi-agent systems), a novel system that enhances transparency and controllability in OMAS through human-in-the-loop planning. 
\tool combines natural language interaction with a graph-based interface that represents plans as editable workflows in a visual programming environment. Through direct manipulation~\cite{1654471}, users can inspect and modify the plan structure--including agent assignments, data flow, and execution order--by interacting directly with nodes and edges in the plan graph.
Additionally, users can invoke LLM assistance to suggest completions, resolve issues, or fill in missing details.
This mixed-initiative~\cite{10.1145/302979.303030} model enables flexible, collaborative planning, combining human insight and expertise with LLM-driven reasoning to iteratively build and refine executable plans.
Our contributions are as follows:
\vspace{-0.3em}
\begin{itemize}[leftmargin=1.5em]
\setlength{\itemsep}{-.3\baselineskip}
    \item \tool, a novel system combining conversational and graph interfaces, providing fine-grained plan exploration and control.
    \item A mixed-initiative planning approach enabling human-LLM collaboration for plan construction and refinement for OMAS.
    \item Experiments and a pilot study demonstrating how \tool improves transparency and controllability in LLM-based planning.
\end{itemize}

\section{\tool System}
\label{sec:system}
\subsection{System Overview}
\label{sec:overview}

\begin{figure}[t]
  \centering
  \includegraphics[width=\columnwidth]{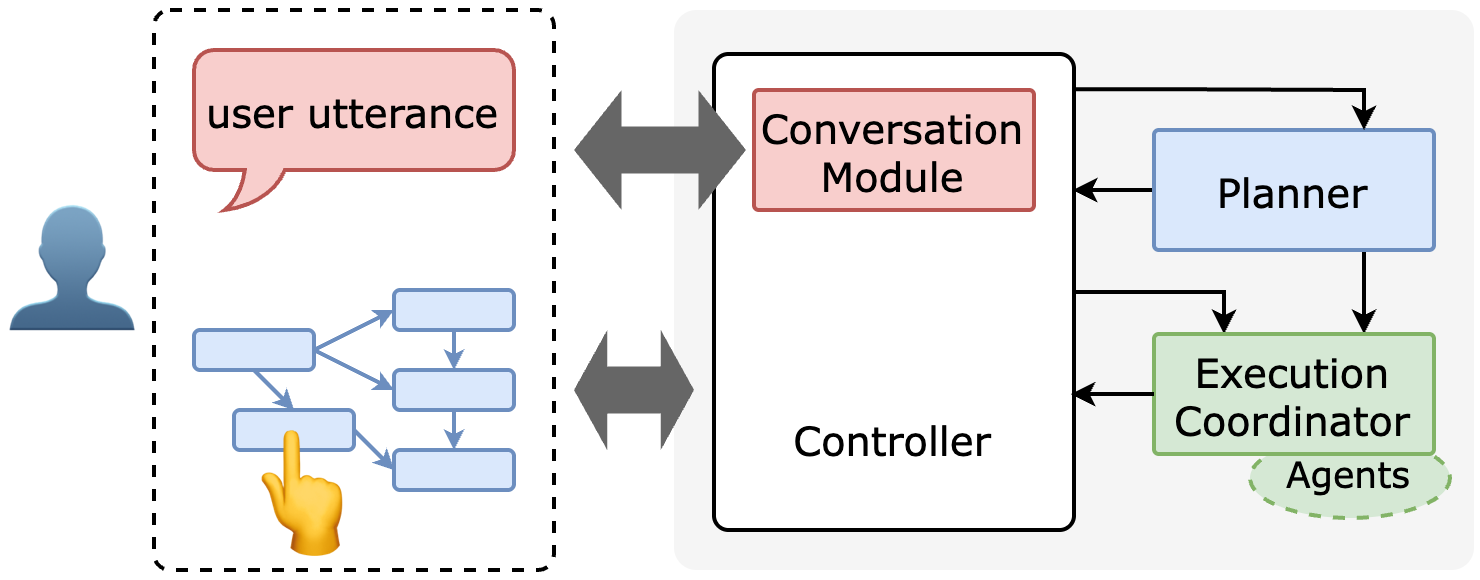}
  \caption{System overview. \tool supports human-in-the-loop planning through natural language interaction and direct manipulation on a plan graph.}
  \label{fig:overview}
  \vspace{-1em}
\end{figure}

\tool consists of four key modules (Fig.~\ref{fig:overview}): a planning module responsible for converting user request into logical plans (\S~\ref{sec:planner}), a conversation module that interprets user utterances and extract intent, an execution coordinator that manages subtask dispatch across agents, and a controller that orchestrates communication between them. 

Users interact with \tool through a dual-panel interface that combines a chat panel (\S~\ref{sec:chatpanel}) for natural language interaction and a plan panel (\S~\ref{sec:planpanel}) for exploring and editing the plan iteratively.
The controller translates user inputs (both natural language feedback and graph edits) into system-level operations that update the plan and coordinate execution. 

Implementation details are listed in Appendix~\ref{app:implementation}.

\paragraph{Plan Model}
A plan is a structured workflow of subtasks and dependencies, represented as a directed acyclic graph (DAG) \cite{zhuo2024rolesllmsplanningembedding, zhang2025planovergraphparallelablellmagent}. Each node in the graph corresponds to a subtask assigned to an agent, specifying its task description, assigned agent, expected inputs, and outputs. Edges define data dependencies from outputs from one node to inputs of another, thereby establishing execution order and information flow. 

This plan representation differs from some prior work, which models plans as node-level DAGs without explicit data mappings or as linear sequences of subtask descriptions. In contrast, our setting requires \textit{coordinating multiple external agents with defined input/output interfaces}, making it essential to \textit{track how outputs of one step connect to inputs of the next}. The DAG structure supports this fine-grained dependency modeling and enables reliable multi-agent execution.

\paragraph{Agents}
Agents (which can be LLM-based, built on top of proprietary models or APIs, or rely on simple tools and function calling) available to the system are described in an agent registry, which defines their names, capabilities, and input/output specifications. This registry serves as a shared source of truth for both the planner
and the execution coordinator.

\subsection{LLM-based Planner}
\label{sec:planner}
\tool uses an LLM to generate and refine plans in an \textit{agent-aware} manner. The planner constructs plans based on agent capabilities and input/output requirements defined in the agent registry.

\subsubsection{Plan Generation}
Plan generation is triggered whenever the conversation module identifies a new user query, representing the user’s latest intent.
This query is passed to the planner along with the agent registry. The LLM planner is prompted to generate a structured, executable plan that decomposes the user's goal into subtasks, assigns each subtask to an appropriate agent, and defines dependencies between them.

\subsubsection{Plan Refinement via User Feedback}
After a plan is generated, users can refine it either through natural language (NL) feedback or through direct manipulation on plan graphs. 

\vspace{-.5em}
\begin{enumerate}[leftmargin=1.1em]
\setlength{\itemsep}{-.3\baselineskip}
    \item \textbf{NL Feedback} Users can provide textual feedback. The planner is then re-prompted with the current plan state, the agent registry, and the user's feedback to produce an updated plan.
    \item \textbf{Direct Manipulation} Alternatively, users can directly edit the plan graph by adding or deleting nodes or edges, modifying task descriptions, reassigning agents, adjusting input/output fields, or updating agent configurations. These changes are immediately reflected in the plan. 
    \item \textbf{LLM Fix} Users may invoke LLM assistance after making partial edits, prompting the planner to complete, validate, or fix the current plan.
\end{enumerate}
\vspace{-.5em}

We posit that NL feedback is well-suited for high-level guidance, such as shaping the overall structure or intent of the plan. In contrast, direct manipulation is more effective for precise or localized adjustments where users aim to retain most of the existing plan. This \textit{mixed-initiative} workflow supports flexible and efficient human-LLM collaboration, leveraging the complementary strengths of NL interaction and structured editing.

\subsection{Interface}
\label{sec:interface}
\tool provide a dual-panel interface that supports both natural language interaction and direct manipulation of a structured plan. This layout enables users to switch fluidly between conversational input and direct edit, supporting a mixed-initiative workflow for human-LLM collaborative planning.

\subsubsection{Plan Panel}
\label{sec:planpanel}
The plan panel (Fig.~\ref{fig:teaser}(B)) displays the generated plan as a directed graph, with the current user query shown in the top-left corner. The plan is visualized as a node-link diagram, where each node represents a task and edges represent data dependencies.

Each node is rendered as a card containing subtask details, including the assigned agent, task description, input/output fields, and execution status. Once a task is executed, its output is shown at the bottom of the node card. Edges are rendered as directional arrows connecting output fields of one node to input fields of another, making data flow across the plan explicit. A green button inside each node card triggers single node execution.

The plan is fully editable via direct manipulation. Users can add new nodes using the “Add Node” button and create edges by dragging from an output to a compatible input. Nodes and edges can be removed by selecting and pressing the delete key. Subtask details (e.g., task descriptions, assigned agents, agent configurations, and input/output variables) can be modified directly within each node card. Task nodes can also be re-positioned freely to improve plan layout. Additionally, intermediate outputs can be manually edited without modifying the plan structure, allowing downstream subtasks to be re-executed with custom inputs.

Control buttons in the top-right corner allow users to execute the entire plan (Execute All), generate a new plan for the current query (Re-plan), or request LLM assistance to complete or fix the current plan (Help).

\subsubsection{Chat Panel}
\label{sec:chatpanel}
The chat panel (Fig.~\ref{fig:teaser}(A)) provides a conversational interface where users interact with the system using natural language. It supports a range of high-level inputs, such as initializing a new plan, modifying the current query, refining an existing plan, or triggering execution.

User messages and system responses are displayed as chat bubbles, forming a clear and traceable interaction history. When a new plan is generated, an execution is triggered, or a plan is refined, the system not only updates the plan panel but also  responds with natural language explanations in the chat panel. This conversational interface complements the plan panel by enabling users to steering the planning process using high-level language, while simultaneously observing plan updates and execution results in context.

\section{Usage Scenarios}
\label{sec:use_case}
\begin{figure}[t]
\centering
 \includegraphics[width=\columnwidth]{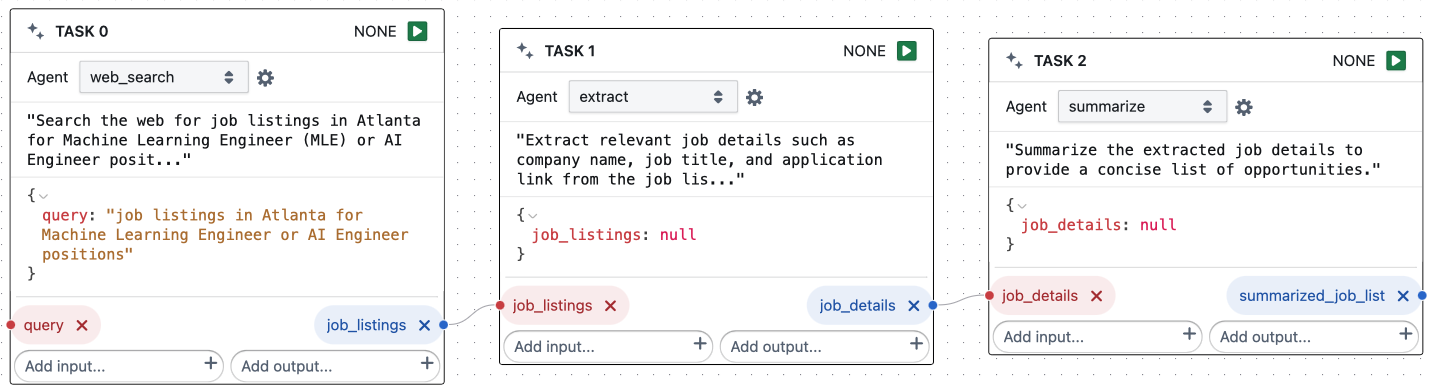}
 \includegraphics[width=.7\columnwidth]{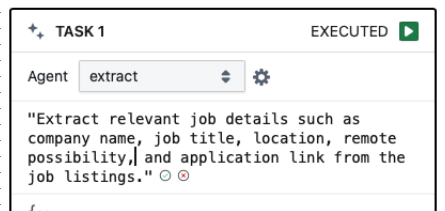}
 \caption{Initial plan generated for job search example (top) and editing task description (bottom).}
 \label{fig:jobsearch}
\end{figure}

\subsection{Searching for a Job}
\label{sec:usecase-job}

Misty is seeking MLE or AI engineering roles in Atlanta. She begins with a query: “Help me find a job in Atlanta. I’m looking for MLE or AI eng positions.” \tool responds with a three-step plan using a web search agent, an extract agent, and a summarization agent (Fig.~\ref{fig:jobsearch}, top).
Misty executes the plan. Intermediate outputs appear in each node, allowing her to observe they contribute to the final answer. When she notices that the search agent returns only five postings, she adjust the agent settings to return 10 results and re-execute the plan.
Next, Misty edits the extract agent's task to include location and remote possibility (Fig.~\ref{fig:jobsearch}, bottom), then re-runs only the modified node and its dependents. Noticing some jobs are outside Atlanta (e.g., Tesla, Palo Alto, is included in Node 1's execution result), she provides feedback via chat: “Filter out jobs that are not in Atlanta” (Fig.~\ref{fig:teaser}). \tool updates the plan by inserting a filtering step.
After re-running the updated plan, Misty is satisfied with the results and proceeds to explore the application links. This scenario highlights \tool’s support for iterative refinement, transparent execution, and granular control. 

\subsection{Solving a Math Problem}
\label{sec:usecase-math}

Brock tries solve a math word problem involving full-priced and discounted glasses (see screenshots in Appendix~\ref{app:screenshots}). The initial plan fails to compute the number of each type, leading to an incomplete solution.
To fix this, Brock adds a placeholder node with a natural language description and two expected outputs, then clicks the Help button to invoke LLM assistance (Fig.~\ref{fig:usecase-math}, top).
During execution of fixed plan, a multiply node produces an incorrect result: multiplying the cost per glass by 60 instead of interpreting it as 60\%. Brock replaces the node with an LLM-based multiply agent to handle the percentage correctly and adds a missing edge to fix a data dependency (Fig.~\ref{fig:usecase-math}, middle).
After these edits, the updated plan successfully solves the problem (Fig.~\ref{fig:usecase-math}, bottom). This example shows how \tool supports plan repair, agent substitution, and user-driven debugging in a mixed-initiative way.

\section{Evaluation}
\label{sec:experiments}
We conduct a quantitative experiment to evaluate plan refinement performance, alongside a pilot study that compares different plan representation formats and feedback modalities.

\subsection{Experiment Setting}
\label{sec:exp_setting}
\begin{table*}[htb!]
    \centering
    \scriptsize  
    \setlength{\tabcolsep}{2.5pt} 
    
    \adjustbox{max width=\linewidth}{
    \begin{tabular}{l|l|l|ccc|ccc|ccc|ccc|ccc|ccc}
        \toprule
        \multirow{2}{*}{\textbf{Model}} 
        & \multirow{2}{*}{\textbf{Dataset}} 
        & \multirow{2}{*}{\textbf{Feedback}} 
        & \multicolumn{3}{c|}{\textbf{Add Node}} 
        & \multicolumn{3}{c|}{\textbf{Remove Node}} 
        & \multicolumn{3}{c|}{\textbf{Add Edge}} 
        & \multicolumn{3}{c|}{\textbf{Remove Edge}} 
        & \multicolumn{3}{c|}{\textbf{Modify (Agent)}} 
        & \multicolumn{3}{c}{\textbf{Modify (I/O)}} \\
        
        \cmidrule(lr){4-6} \cmidrule(lr){7-9} \cmidrule(lr){10-12} 
        \cmidrule(lr){13-15} \cmidrule(lr){16-18} \cmidrule(lr){19-21}
        & & & \textbf{Acc} & \textbf{ISO} & \textbf{GED}
        & \textbf{Acc} & \textbf{ISO} & \textbf{GED}
        & \textbf{Acc} & \textbf{ISO} & \textbf{GED}
        & \textbf{Acc} & \textbf{ISO} & \textbf{GED}
        & \textbf{Acc} & \textbf{ISO} & \textbf{GED}
        & \textbf{Acc} & \textbf{ISO} & \textbf{GED} \\
        
        \midrule
        
        \multirow{6}{*}{\textbf{GPT-4o}}
        & \multirow{3}{*}{\textbf{GSM8K}}
        & \textbf{Detailed}  & \underline{70.97} & 96.77 & 0.05   & 96.77 & 93.50 & 0.26   & 93.55 & 100   & 0.00   & 93.55 & 100   & 0.00   & 95.16 & 100   & 0.00   & \underline{98.38} & 93.54 & 0.19 \\
        & & \textbf{Vague}     & 67.74 & 82.25  & 0.73   & 90.32 & 88.71 & 0.44   & 95.16 & 95.16 & 0.08   & 83.9  & 91.9  & 0.13   & 93.5  & 87.1 & 0.52   & 95.16 & 90.3  & 0.35 \\
        & & \textbf{DM + Fix}  & \textbf{93.54} & 90.32 & 0.31   & 100    & 100    & 0     & 100    & 100    & 0    & 100    & 100    & 0    & 100    & 100    & 0       & \textbf{96.77} & 91.94 & 0.29 \\
\cmidrule(lr){2-21}
        & \multirow{3}{*}{\textbf{Multi-step}}
        & \textbf{Detailed}  & \underline{19.6}  & 95.2  & 0.07   & 52    & 68.4  & 0.79   & 81.2  & 100   & 0.00   & 80.8  & 99.6  & 0.01   & 74    & 93.2   & 0.07   & \underline{60}    & 82.4  & 0.21 \\
        & & \textbf{Vague}     & 6.4   & 27.2    & 1.83  & 41.6  & 53.2  & 1.71   & 74    & 96.4  & 0.07   & 79.6  & 97.6  & 0.02   & 54    & 65.6    & 1.58   & 38.4  & 88    & 0.16 \\
        & & \textbf{DM + Fix}  & \textbf{11.2}  & 43.8  & 4.42   & 100    & 100    & 0       & 100    & 100    & 0      & 100    & 100    & 0      & 100    & 100    & 0       & \textbf{37.2}  & 50.6  & 4.28 \\
        
        \midrule

        \multirow{6}{.8cm}{\textbf{Llama-\\3.3-70B}}
        & \multirow{3}{*}{\textbf{GSM8K}}
        & \textbf{Detailed}  & \underline{72.58} & 90.32 & 0.58 & 51.61 & 95.16 & 0.13 & 75.81 & 95.16 & 0.10 & 74.19  & 93.55 & 0.18 & 74.19 & 93.55 & 0.12 & \underline{71.77} & 92.74 & 0.29 \\
        & & \textbf{Vague}     & 64.51 & 85.48 & 0.53 & 50.0 & 61.29 & 1.29 & 74.19 & 91.94 & 0.19 & 72.58 & 88.71 & 0.39 & 70.97 & 82.26 & 1.08 & 66.94 & 83.87 & 0.85  \\
        & & \textbf{DM + Fix}  & \textbf{77.05} & 65.00 & 1.87 & 100 & 100 & 0 & 100 & 100 & 0 & 100 & 100 & 0 & 100 & 100 & 0 & \textbf{90.32}  & 83.33 & 0.83 \\
\cmidrule(lr){2-21}
        & \multirow{3}{*}{\textbf{Multi-step}}
        & \textbf{Detailed}  & \underline{5.60} & 8.50 & 8.33 & 19.60 & 72.43 & 0.59 & 64.40 & 85.20 & 1.14 & 65.60 & 88.28 & 0.80 & 37.6 & 74.13 & 2.08 & \underline{33.0} & 66.60 & 2.62 \\
        & & \textbf{Vague}     & 9.20 & 13.97 & 5.69 & 0.40 & 1.61 & 8.63 & 59.60 & 83.60 & 1.32 & 18.0 & 24.24 & 6.69 & 33.10 & 71.06 & 2.59 & 13.30 & 20.36 & 10.44 \\
        & & \textbf{DM + Fix}  & \textbf{12.6} & 23.41 & 6.69 & 100 & 100 & 0 & 100 & 100 & 0 & 100 & 100 & 0 & 100 & 100 & 0 & \textbf{26.87} & 25.20 & 7.02 \\

        \bottomrule
    \end{tabular}
    }
    \centering
    \caption{Plan refinement performance across operation types for different feedback formats and models. Metrics include execution accuracy (Acc ↑), isomorphic subgraph match (ISO ↑), and graph edit distance (GED ↓). Highlighted are the \textbf{DM+Fix} and baseline \underline{Detailed Feedback} performance for complex operations.}
    \vspace{-1em}
    \label{tab:compare}
\end{table*}

\paragraph{Datasets and Tasks}
Our experiments utilize two datasets focused on math reasoning: GSM8K (grade-school-level word problems, \citealp{cobbe2021training}) and Multi-Step Arithmetic from BIG-Bench Hard (complex equation-format problems, \citealp{suzgun2022challenging}). We select math problems because their solutions have limited variability in the correct plan structure, unlike other tasks that may have multiple correct approaches involving different sets of agents, making them easier to evaluate.

For each dataset, we randomly sample 50 tasks and manually generate a correct plan $p_1$, which is then validated by the authors.
We then \textit{artificially modify} each correct plan by randomly applying one operation (e.g., adding or removing a node or edge, or altering a subtask specification) to produce an incorrect version $p_0$.
We assess the planner's ability to refine $p_0$ back to the correct plan $p_1$ using three kinds of feedback formats: detailed natural language feedback, vague/underspecified natural language feedback, and partial manipulation with LLM assistance.
After the planner generates a refined plan $p'_1$, we compare it to the original correct plan $p_1$.
The list of modification operations and example feedbacks are included in Appendix~\ref{app:exp_modification_ops}.

\paragraph{Metrics}
We evaluate refined plans using the execution accuracy of refined plans and graph similarity to the original correct plans. These metrics capture functional correctness (whether the task is solved) and structural correctness (alignment with the original plan).
For graph similarity, we employ: (1) isomorphism (ISO), which measures whether the graphs are structurally identical with matching agent assignments;
(2) graph edit distance (GED), the minimum number of edit operations required to transform one graph into the other.

\subsection{Experiment Results}

Table~\ref{tab:compare} compares the effectiveness of feedback formats across plan refinement operations using the GPT-4o and Llama-3.3-70B-Instruct models.

Compared to vague NL feedback, detailed NL feedback achieves higher performance across nearly all refinement operations, confirming that precise and explicit instructions enable the LLM planner to reliably recover correct plans.
However, this assumes that users are both able and willing to articulate details, which can impose cognitive burden, especially in complex or unfamiliar domains.
Vague NL feedback, by contrast, is less effective: its ambiguity limits the planner's ability to accurately infer users' refinement intent.
These results highlight that while natural language interactions are useful, their effectiveness depends heavily on the specificity of user input. As a result, they cannot be solely relied upon for plan refinement, especially when user intent is implicit, ambiguous, or difficult to express in language. 

Direct manipulation with LLM assistance (DM+Fix) offers a practical alternative, allowing users to make partial edits on plan while relying on the LLM to complete and correct the plan. For simple, single-step operations (e.g., remove node, add or remove edge, and modify agent assignment), direct manipulation alone achieves near-perfect accuracy. For more complex operations that involve multiple interdependent changes (e.g., adding a new node and connecting its dependencies), DM+Fix outperforms vague feedback and performs comparably to detailed feedback, while requiring less user effort. 

We also observe lower performance on the Multi-Step Arithmetic dataset due to the complexity of its generated plans. Multi-step plans require intricate output-input dependencies between nodes. Modification operations can easily disrupt these links, making accurate refinement challenging.

Overall, GPT-4o consistently outperforms Llama-3.3-70B, often by a significant margin. However, both models exhibit similar performance trends, indicating comparable behavior despite differences in absolute metrics.

\begin{table*}[tb]
\centering
\footnotesize
\begin{tabular}{l|ccc|cc}
\toprule
\multicolumn{1}{p{2.4cm}|}{\centering \textbf{Phase}\\(Plan$\rightarrow$ Feedback)} & \multicolumn{1}{p{1.5cm}}{\centering \textbf{Completion\\Time (sec)}} & \multicolumn{1}{p{0.6cm}}{\centering \textbf{Word\\Count}} & \multicolumn{1}{p{1.5cm}}{\centering \textbf{Interaction\\Count}} & \multicolumn{1}{|p{1.5cm}}{\centering \textbf{False\\Acceptance}} & \multicolumn{1}{p{2cm}}{\centering \textbf{Post-Feedback\\Accuracy}}\\
\midrule
\circled{1} Text$\rightarrow$\blackcircled{1} Text & 173.72 & 18.09 & - & 22.22\% & 80.56\%\\
\circled{2} Graph$\rightarrow$\blackcircled{1} Text & 149.98 & 12.39 & - & 11.11\% & 86.11\%\\
\circled{2} Graph$\rightarrow$\blackcircled{2} DM & 155.37 & - & 2.16 & 0\% & 88.89\%\\
\bottomrule
\end{tabular}
\caption{User study results across phases where participants were presented with plans (textual or graph) and provided feedback (text or direct manipulation of graph). Average task completion time (seconds), textual feedback word count, direct manipulation interaction count, false acceptance rate, and post-feedback accuracy are reported.}
\vspace{-1em}
\label{tab:study_results}
\end{table*}

\subsection{Pilot Study}

We conducted a small-scale pilot study to explore how users provide feedback to refine LLM-generated plans. The study compared plan representation formats (\circled{1} text vs. \circled{2} graph) and feedback modalities (\blackcircled{1} textual comments vs. \blackcircled{2} partial graph edits). Participants were presented with flawed plans and provided feedback across three phases combining these conditions. Detailed study design and results are provided in Appendix~\ref{app:user_study}.

Table 2 shows that execution accuracy of refined plans is higher when the original plan is presented as a graph rather than text, and when feedback is given via direct manipulation (followed by LLM fix assistance) rather than natural language. Additionally, participants completed tasks faster and provided more concise textual feedback when plans were presented in graph format compared to textual plans. Participants also accepted incorrect plans as correct twice as often when working with textual plans. These findings align with survey results in which users found graph presentations easier to understand and debug and preferred graph editing over textual feedback (see Appendix~\ref{app:exit_survey_results}). These results highlight the benefits of our graph visualization for transparency and interpretability over conventional chat-based agentic systems.

\section{Related Works}
\label{sec:relwork}
\paragraph{Multi-Agent Systems}
Our work builds on recent trends in multiple specialized agents AI systems, referred to as multi-agent systems (MAS), compound AI, agentic workflows, AI pipelines, etc. While traditional MAS emphasize agent autonomy, cooperation, and distributed decision-making~\cite{wooldridge2009introduction,Stone2000}, our focus is on a class of \textit{centrally orchestrated} systems that coordinate pre-defined agents, i.e., each implementing a modular function or service. 
In this \textit{orchestrated} MAS setting, agents are not proactive or autonomous; instead, they execute assigned tasks upon request. This design aligns with recent notions of compound AI~\cite{kandogan2024blueprintarchitecturecompoundai} and agentic workflows~\cite{qiao2025benchmarking}.

\paragraph{LLM-Based Planning}
LLM-based planning has gained popularity due to language models’ ability to reason step-by-step and decompose tasks without domain-specific training~\cite{valmeekam2023on,huang2024understandingplanningllmagents}. Many systems interleave planning and execution, generating and executing one step at a time based on observed outcomes~\cite{yao2023react,schick2023toolformer,prasad2023adapt,wang2023voyageropenendedembodiedagent}. This paradigm
enables flexible adaptation but lacks a global, inspectable plan structure. In contrast, our system adopts a plan-then-execute approach: it generates a complete multi-agent plan upfront, enabling granular inspection and refinement by humans.

\paragraph{Interactive AI Workflow Systems}
Our system shares common goals with LLM chains / ML workflow systems~\cite{wu2022aichains,promptsapper,arawjo2024chainforge,jigsaw},
which offer visual programming interfaces for assembling modular components into executable chains or ML pipelines. While these systems are conceptually similar to plans in OMAS, they typically require users to manually construct plans from scratch or rely on predefined templates.

InstructPipe~\cite{zhou2025instructpipe}, ChainBuddy~\cite{chainbuddy}, and Low-code LLM~\cite{cai-etal-2024-low-code} use LLMs to generate structured pipelines/workflows from NL descriptions. While our interface shares similarities in combining NL with visual interactions, our work targets OMAS, where planning must be agent-aware, with explicit data flow across agents. Unlike systems focused on initial generation, \tool supports mixed-initiative refinement, allowing users to collaboratively build and update plans.

\section{Conclusion}
\label{sec:conclusion}
We presented \tool, a system addressing key limitations in current LLM-based planning for orchestrated multi-agent systems. By combining conversational and graph-based interfaces, \tool enhances transparency and controllability through flexible human-in-the-loop collaboration. Preliminary results demonstrate its effectiveness in interactive plan refinement.

Future work includes applying \tool to high-stakes domains like healthcare and finance, where precise and controllable planning by domain experts is essential. We plan to expand user interactions beyond basic graph edits to support operations such as freezing, merging, splitting, replacing tasks, and enforcing structural constraints (e.g., “A must precede B, but not coincide with C”). To improve scalability, we aim to enhance LLM assistance in verifying plans and executions, enabling users to focus on ambiguous or problematic areas~\cite{sung2025verilahumancenteredevaluationframework}. Finally, real-world deployments and user studies will help us assess which interactions users prefer, how they impact trust, and how to further refine the system for practical use.

\section*{Ethics Statement}
We promote the collaboration of LLM-based planners and humans, which can be beneficial for various tasks. It is important to take note of the responsible use of such systems. Over-reliance on LLM-based planners may expose inherent biases present within such models which could influence decision-making in real-world scenarios. 
Furthermore, it is important for humans to be accountable of their actions, that is, bad actors may exploit such systems to refine plans to suit their own benefits and biases. 

We also emphasize the need for privacy and data protection. If the planner handles personal or sensitive data, human intervention may introduce privacy risks. Such issues may be mitigated by using role-based access to such systems as well as data anonymization. 

As LLMs continue to be utilized for planning, it is important to do so with responsible human monitoring which ensures planning and decision-making is transparent, accountable and unbiased.



\bibliography{anthology,custom}

\appendix
\newpage
\section{Implementation Details}
\label{app:implementation}
\tool is built using a React frontend and a Python backend with FastAPI for communication. The planner and conversation module are powered by OpenAI's GPT-4o.
To simulate an OMAS setting, we created specialized agents using Python functions, APIs, and LLMs, based on a curated subtask taxonomy.\footnote{While a full OMAS system would involve explicit models, tools, and predefined agents, simulating the agents in this way simplifies the setup and allows us to demonstrate the core interactive planning functionality. \tool can be easily incorporated with actual agents by adding them to the agent registry and wiring them into the execution flow.}

\subsection{List of Prompts}
\label{app:prompts}

We provide the prompts used by our planner and conversation module, with the system prompt listed at the top and the user prompt at the bottom.

\begin{promptbox}{Prompt used for planing} \label{app:planning_prompt}
\scriptsize
You are a planner responsible for creating high-level plans to solve any tasks using a set of agents.\\
Your goal is to break down a given task into a sequence of subtasks that, when executed correctly by the appropriate agents, will lead to the correct solution.\\
A plan should have at least 2 steps.\\
\\
For each step in the plan:\\
1. Describe the subtask the agent must perform.\\
2. Provide a brief, self-contained description of the expected inputs and outputs. Do not include any specific values or examples.\\
3. Generate an instruction prompt for the agent.\\
\\
Represent your plan as a graph where each node corresponds to a step, and each edge represents a dependency between two steps i.e., a step's output is used as an input for a subsequent step.\\
If a node requires the output from a previous node as an input, ensure it is included in the edge list.\\
An input variable for a node represented is a tuple, where the first item is an input description, the second item is the value of the variable if it can be predetermined without executing the plan.\\
If is dependent upon preceding nodes, set null. DO NOT INFER THE VALUE. DO NOT EXECUTE THE STEPS.\\
The output should be structured in the following JSON format:\\
$\{$\\
    'nodes': <list of JSON nodes {{'id': <node id as integer>, 'name': <assigned agent name>, 'task': <instruction prompt>, 'input': <list of tuple (input var, its value)>, 'output': <list of outputs>}}>,\\
    'edges': <list of JSON edges {{'src\_node': <source node id>, 'dest\_node': <destination node id>, 'src\_output': <output variable name>, 'dest\_input': <input variable name>}}>\\
$\}$\\
\\
eg.\\
\{\texttt{plan demonstration examples}\}\\
\\
Here are the available agents:\\
\texttt{\textasciigrave\textasciigrave\textasciigrave}\\
\{\texttt{agent registry}\}\\
\texttt{\textasciigrave\textasciigrave\textasciigrave}\\
\\
For identify\_operands, ensure you repeat the query in the task. Sometimes, the query may require a multiplier eg. "..twice of", divisor eg. "divide by x", percentage, in a later task. Ensure all such operations are also captured in identify\_operands.\\
There may be multiple inputs from one node to another. In that case, ensure you define separate edges from one node to the other.\\
For some agents, ensure that input order is correct, e.g., when calculating profit, revenue - cost is different from cost - revenue. so input should be [revenue, cost] order.\\

\tcblower
\scriptsize
\{\texttt{task query}\}
\end{promptbox}

\begin{promptbox}{Prompt used for refining plan}\label{app:refinement_prompt}
\scriptsize
<same system prompt as planning>

\tcblower
\scriptsize
Given the original plan, refine it according to user feedback\\
\\
Original Plan:\\
\{\texttt{prev plan}\}\\
\\
User Feedback:\\
\{\texttt{feedback}\}
\end{promptbox}

\begin{promptbox}{Prompt used for completing/fixing plan}\label{app:fix_prompt}
\scriptsize
<same system prompt as planning>

\tcblower
\scriptsize
Given a query, an initial plan will be given to you. The initial plan may be incomplete or incorrect.\\
Your job is to complete or fix the plan. Stay as true to the initial plan as you can.\\
\\
Query:\\
\{\texttt{query}\}\\
\\
Intial Plan:\\
\{\texttt{plan}\}\\
\end{promptbox}

\begin{promptbox}{Prompt used for response generation}\label{app:response_prompt}
\scriptsize
You are a natural language interface for a multi-agent system.\\
This system creates a plan to answer a user query and executes it using AI agents.\\
Your task is to explain the actions triggered by the user input and clearly communicate the system's output in a very short (max 1-2 line) response.\\
Do not mention anything else. Write down only plain text.

\tcblower
\scriptsize
\textbf{Case 1: Response after new plan generation}\\
Generate a very short (max 1-2 line) response to a user query to generate a plan.
The response should simply provide a high level response of what the plan does, and minor details such as number of steps. 
\\
User Query: \{\texttt{query}\}\\
Plan: \{\texttt{plan}\}\\
\\
\textbf{Case 2: Response after plan execution}\\
Generate a very short (max 1-2 line) response to a user query to execute a plan or a single node.\\
The response should simply provide a high level response of the execution, and minor details such as final result. \\
\\
User Query: \{\texttt{query}\}\\
Plan: \{\texttt{plan}\}\\
\\
\textbf{Case 3: Response after feedback-based refinement}\\
Generate a very short (max 1-2 line) response to a user query to interact with a plan.\\
The response should simply provide a high level response of the plan which was interacted with and what change took place.\\
\\
Interaction Type: \{\texttt{interaction}\}\\
Plan: \{\texttt{plan}\}
\end{promptbox}

\section{Modification and Feedback Templates}
\label{app:exp_modification_ops}

\begin{table*}[htb!]
    \centering
    \scriptsize
    \setlength{\tabcolsep}{2.5pt}
    
    \begin{tabular}{l l p{5.2cm} p{3cm} p{3cm}}
        \toprule
        \textbf{Modification} & \textbf{Refinement Ops.}
        & \textbf{Detailed NL Feedback} 
        & \textbf{Vague NL Feedback} 
        & \textbf{Direct Manipultaion} \\
        \midrule
        
        Remove arbitrary node    
        & \textbf{Add Node}    
        & Add a \texttt{\{agent\}} node connecting \texttt{\{prev\}} to \texttt{\{next\}}
        & Add a \texttt{\{agent\}} node 
        & Add \texttt{agent} node + Fix \\
        
        Add arbitrary node        
        & \textbf{Remove Node}  
        & Remove a superfluous \texttt{\{agent\}} node  
        & Remove a superfluous node 
        & Remove a specific node \\
        
        Delete arbitrary edge     
        & \textbf{Add Edge}     
        & Add an edge between \texttt{\{source id\}} and \texttt{\{target id\}}  
        & Add a missing edge 
        & Add an edge connecting nodes \\
        
        Add arbitrary edge        
        & \textbf{Remove Edge}  
        & Remove a superfluous edge between \texttt{\{source id\}} and \texttt{\{target id\}}  
        & Remove a superfluous edge 
        & Remove a specific edge \\
        
        Incorrect agent in node
        & \textbf{Modify (Agent)} 
        & Update node \texttt{\{id\}} to have the correct agent  
        & Assign a correct agent 
        & Change the assigned agent \\
        
        Random I/O change
        & \textbf{Modify (I/O)} 
        & Update node \texttt{\{id\}} with valid inputs and outputs  
        & Set valid inputs and outputs 
        & Add/remove an I/O + Fix \\
        
        \bottomrule
    \end{tabular}
    \vspace{-.3em}
    \caption{Modifications performed to test plan refinement capabilities, along with corresponding templates for detailed \& vague natural language feedbacks and direct manipulation with LLM assistance.}
    \vspace{-.5em}
    \label{tab:operations_templates}
\end{table*}

We apply single-step modifications to a correct plan $p_1$ to generate an incorrect plan $p_0$. The planner's task is to refine $p_0$ back to $p_1$ using three types of feedback, as shown in Table \ref{tab:operations_templates}. NL feedback is generated from templates and fed to the planner, whereas direct manipulation feedback is applied via graph edits, followed by LLM-assisted fixes. Although the associated refinement operation is included in the table for clarity, it is not revealed to the planner (i.e., the planner must infer it).

\section{Pilot Study Details}
\label{app:user_study}

\subsection{Study Design}
\paragraph{Participants}
We recruited nine participants from an industry research laboratory, comprising interns, engineers, and research scientists. All participants were proficient in English, based in the United States, held at least a graduate-level degree, and had prior experience working with LLMs. The study objectives and how their input would be used were clearly explained to the participants.

\paragraph{Tasks}
We sampled 12 math word problems from the GSM8K dataset and constructed imperfect plans for each, following Table~\ref{tab:operations_templates}. To control task difficulty, we included both easy and medium tasks: easy tasks were created by applying a single modification to a gold (reference) plan, while medium tasks involved two or more modifications.

In each task, participants were presented with a flawed plan and asked to provide feedback to improve it. Plans were shown in one of two formats: \circled{1} textual descriptions in the chat panel, or \circled{2} graph representations in the plan panel. Participants provided feedback through two modalities: \blackcircled{2} natural language comments and \blackcircled{1} partial graph edits, including adding or deleting nodes or edges and modifying input/output variables. Graph edits were intended as partial signals to guide the planner, rather than complete corrections.

\paragraph{Procedure}
The study followed a within-subjects design, with tasks evenly divided across three phases: (1) participants received textual plans and provided textual feedback (\circled{1}$\rightarrow$\blackcircled{1}); (2) participants received plan graphs and provided textual feedback (\circled{2}$\rightarrow$\blackcircled{1}); and (3) participants received plan graphs and performed partial graph edits to guide the LLM planner (\circled{2}$\rightarrow$\blackcircled{2}). Each phase included four tasks. Both the assignment of tasks to phases and the order of phases were randomized for each participant to control for ordering effects.

Note that participants were not asked to confirm or reject the actual refinements based on their feedback; rather, their input was collected and post-processed to assess whether it led to improvements.

\paragraph{Post-Study Survey}
After completing all tasks, participants answered an exit survey comparing the two plan representation formats (text and graph) and the two feedback modalities (textual feedback and direct manipulation of the graph). The survey assessed ease of understanding and issue detection for plan representations, as well as ease of use, cognitive effort, and preferences for feedback modalities.
The full list of questions are:
\vspace{-.3em}
\begin{itemize}[leftmargin=1.1em]
\setlength{\itemsep}{-.3\baselineskip}
  \item The plan was easy to understand in text format.
  \item The plan was easy to understand in graph format.
  \item It was easy to detect issues or flaws in the text plan.
  \item It was easy to detect issues or flaws in the plan graph.
  \item It was easy to provide useful feedback by writing textual feedback.
  \item It was easy to provide useful feedback by partially editing the plan graph.
  \item Providing feedback by writing textual feedback required a lot of mental effort.
  \item Providing feedback by partially editing the plan graph required a lot of mental effort.
  \item I would prefer to use text feedback for future tasks.
  \item I would prefer to use partial graph editing for future tasks.
\end{itemize}

\begin{figure*}[tbh]
\centering
 \includegraphics[width=\linewidth]{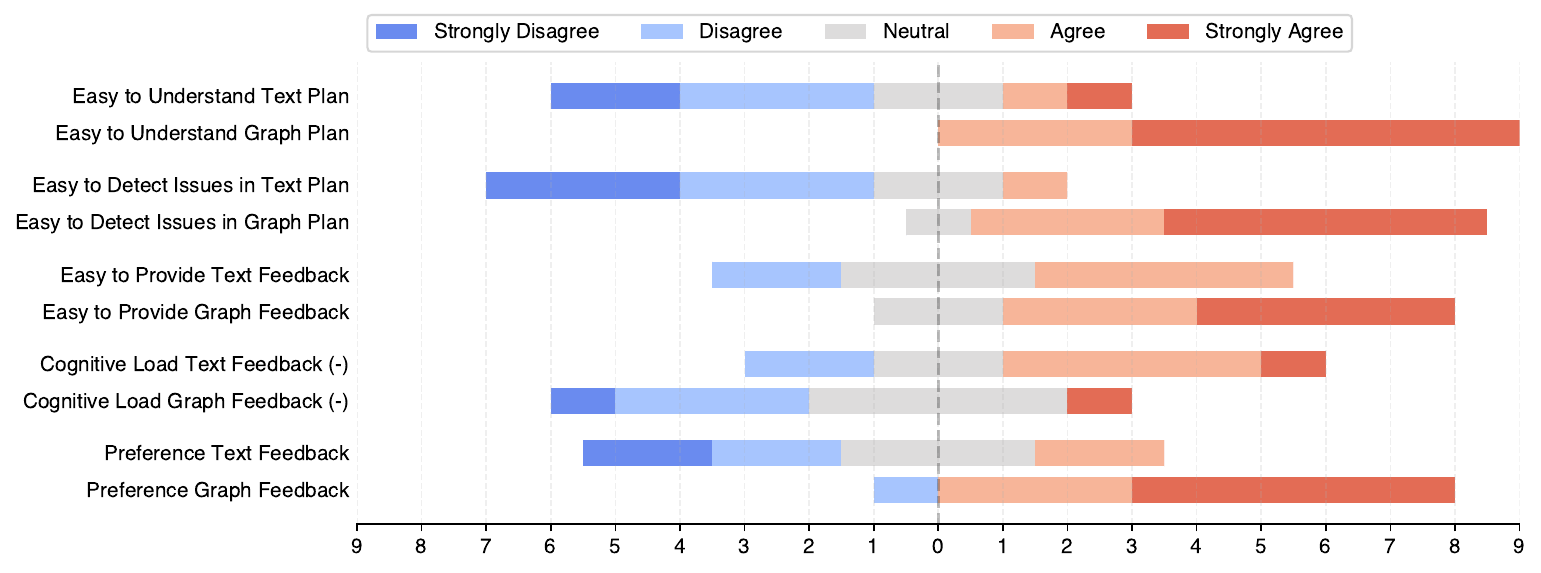}
 \caption{Participant responses from the exit survey, with each row representing 9 responses. Graph-based plan representations were perceived as significantly easier to interpret and debug than textual plans. Participants also preferred partial graph editing over textual feedback, finding it easier to provide and less mentally demanding.}
 \label{fig:exit_survey_ratings}
\end{figure*}

\subsection{Exit Survey Results}
\label{app:exit_survey_results}

Figure~\ref{fig:exit_survey_ratings} summarizes participants' responses from the exit survey, capturing perceived usability and preferences across plan representation formats and feedback modalities.

\paragraph{Plan Representations}
Participants found graph-based plans considerably easier to interpret and debug than textual plans. One participant noted, ``\textit{[...] if graph visualization is provided, issues like missing edges are easy to be detected immediately.}'' This suggests that the visual structure of the graph helped users reason about dependencies and execution flow between agents.

\paragraph{Feedback Modalities}
Participants rated partial graph editing more favorably than textual feedback in terms of ease of use and lower mental effort. Also, participants expressed a strong preference for using graph edits in future tasks. One participant commented, ``\textit{Textual seems more helpful for high-level feedback and graph editing is more suitable for detailed editing [...].}'' These responses indicate that users perceive direct manipulation as a complementary and intuitive addition to conversational feedback for guiding LLM planners.

\section{Additional Screenshots}
\label{app:screenshots}

In this section, we present additional screenshots of our system, captured while following the usage scenario described in \S~\ref{sec:usecase-math}.

\begin{figure*}[t]
\centering
  \includegraphics[width=\linewidth,cframe=gray]{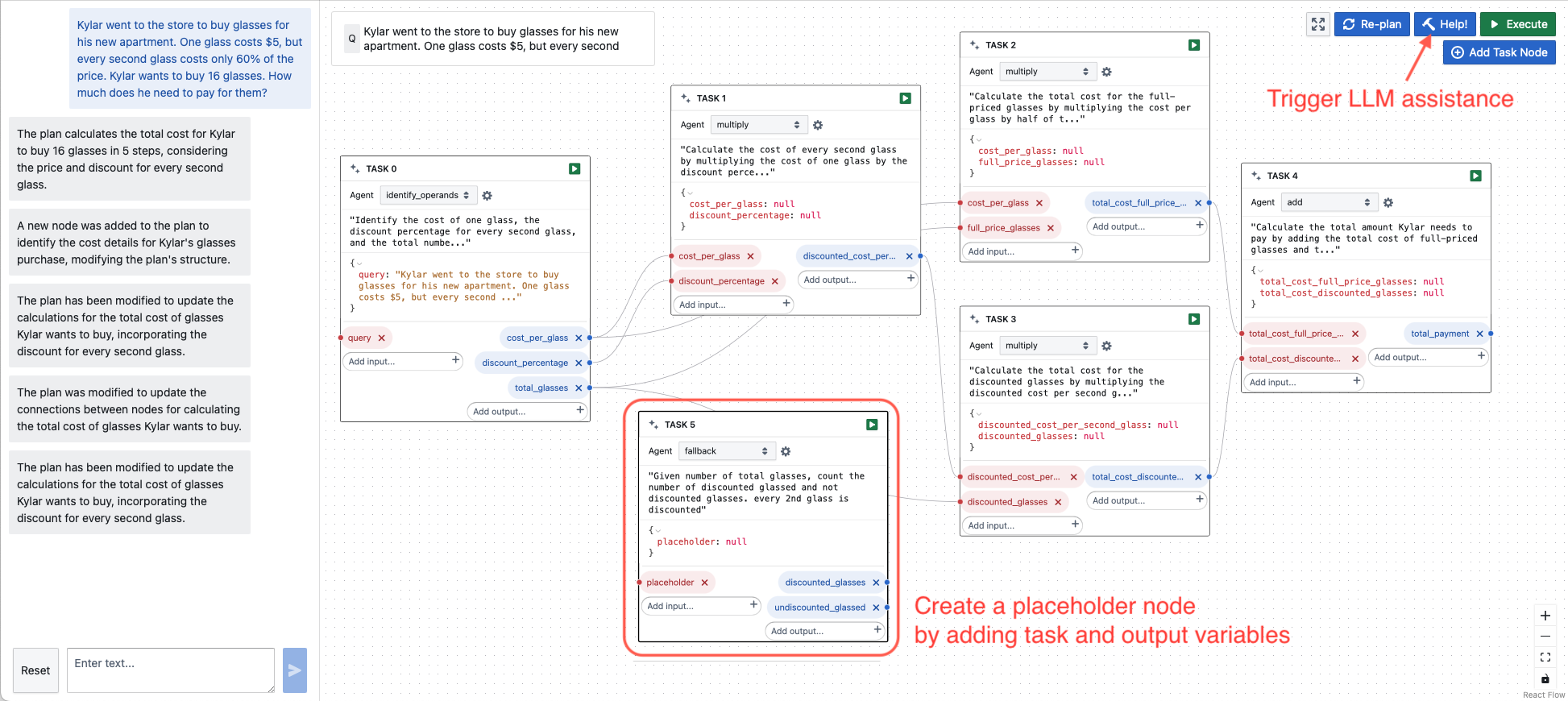}
\par\nointerlineskip\vspace{.5cm}\nointerlineskip
  \includegraphics[width=\linewidth,cframe=gray]{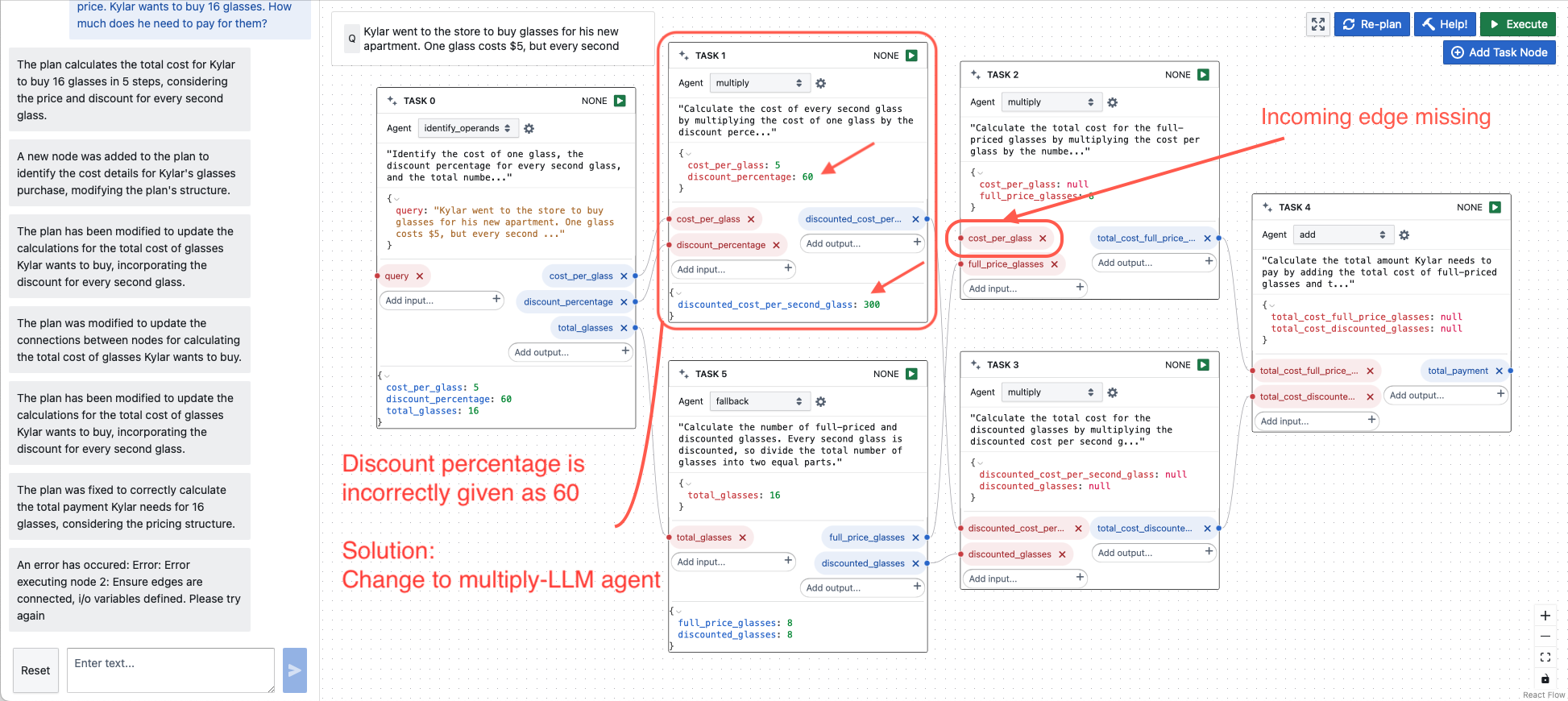}
\par\nointerlineskip\vspace{.5cm}\nointerlineskip
  \includegraphics[width=\linewidth,cframe=gray]{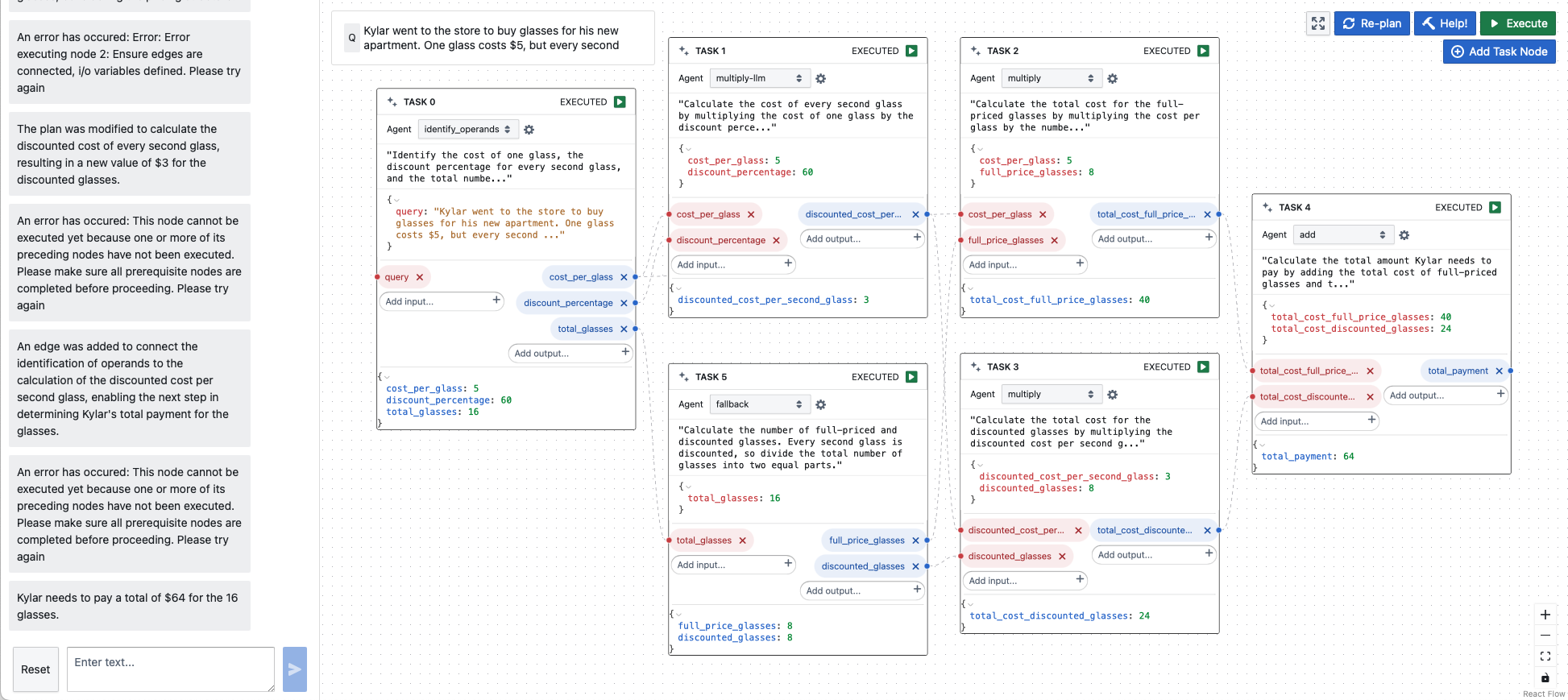}
  \caption {Screenshots providing a walkthrough of the use case scenario described in \S~\ref{sec:usecase-math}.}
  \label{fig:usecase-math}
\end{figure*}

\end{document}